\documentclass[superscriptaddress,12pt]{revtex4-2} 

\usepackage{mathtools}

\usepackage{booktabs,makecell,array}
\newcolumntype{P}[1]{>{\raggedright\arraybackslash}p{#1}}

\usepackage{amsmath}  
\usepackage{amsfonts} 
\usepackage{graphicx} 
\usepackage{esvect}
\usepackage{mathtools}

\usepackage{xcolor}
\usepackage{appendix}
\usepackage{siunitx} 
\usepackage{mathrsfs}

\usepackage{placeins}

\usepackage[caption=false]{subfig} 

\usepackage{pbox}

\usepackage{mathtools}
\makeatletter
\newcommand{\vast}{\bBigg@{3}}
\newcommand{\Vast}{\bBigg@{3.5}}
\makeatother

\begin{document}


\title{Size and Shape of Fuzzy Spheres \\ from Matrix/Membrane Correspondence}


\author{Hai H. Vo}
\thanks{These authors contributed equally.}
\affiliation{Massachusetts Institute
of Technology, \\ 77 Massachusetts Ave., Cambridge, MA 02139, USA}

\author{Olivia M. Markowich}
\thanks{These authors contributed equally.}
\affiliation{Scripps College, Claremont, CA 91711, USA.}

\author{Angeline Hu}
\thanks{These authors contributed equally.}
\affiliation{Claremont McKenna College, Claremont, CA 91711, USA.}

\author{Nguyen H. Nguyen}
\affiliation{Ho Chi Minh City University of Technology, \\ 268 Ly Thuong Kiet, Ho Chi Minh 700000, Vietnam.}

\author{Trung V. Phan}
\email{{tphan@natsci.claremont.edu}}
\affiliation{Department of Natural Sciences, Scripps and Pitzer Colleges, \\ Claremont Colleges Consortium, Claremont, CA 91711, USA}

\begin{abstract}
We study the size and shape statistics of ground state \textit{fuzzy spheres} when projected onto the transverse plane, utilizing the regularized SU$(N=2)$ matrix model in $\mathscr{D}=(1+3)$-dimensional spacetime. We show that they appear as ellipses, from matrix/membrane correspondence. With our numerical and analytical approximation for the ground state wavefunction, we provide estimations for their expectation surface areas, perimeters, eccentricities, and shape-parameters. These geometric constants of quantum membranes deviate drastically from classical mechanics.  
\end{abstract}

\date{\today}

\maketitle 

\section{Introduction}

Membranes (M2-branes) are hypothesized to be among the fundamental extended objects of M-theory \cite{duff1999m}, a conjectured quantum gravity framework unifying all string theories \cite{witten1995string,hull1995unity,duff1999eleven}. A free membrane has an action proportional to the world-volume it sweeps out in $\mathscr{D}$-dimensional spacetime \cite{dirac1962extensible,gnadig1978dirac}. In the light-front formulation, one selects a null coordinate as the evolution parameter (“time”) and imposes the world-volume reparametrization (diffeomorphism) constraints, which then determine the motion along the conjugate null direction. What remains dynamical are only the $(\mathscr{D}-2)$ transverse directions (independent propagating local degrees of freedom). After removing the constant center-of-mass motion and performing quantization (replacing Poisson brackets by Lie algebra commutators), this dynamics reduces to an SU$(N\rightarrow\infty)$ matrix quantum mechanics model \cite{hoppe1982quantum,hoppe1989quantum,banks1999m}. The truncation to a finite value $N$ \cite{hoppe1982quantum} provides a matrix regularization that captures low-resolution membrane fluctuations, similar to how quantum string snapshots
are interpreted (see Fig. 1A) \cite{karliner1988size}. There exists a matrix/membrane correspondence \cite{kabat1997spherical,shimada2004membrane} that maps $(\mathscr{D}-2)$ $N\times N$ matrices to membrane configurations in $(\mathscr{D}-2)$-dimensional transverse space; 
in certain setups, this mapping can even be interpreted as a second-quantized description of multiple membranes \cite{taylor2001m,dasgupta2002introduction}. Although quantum strings in simple spacetime backgrounds are well understood \cite{green1987superstring,green2012superstring,polchinski1998string,polchinski2005string}, quantum membranes \cite{collins1976classical} remain comparatively underexplored, and theoretical tractability continues to be a major challenge \cite{douglas1998matrix,trzetrzelewski2010spiky}.

\begin{figure*}[!htbp]
\includegraphics[width=\textwidth]{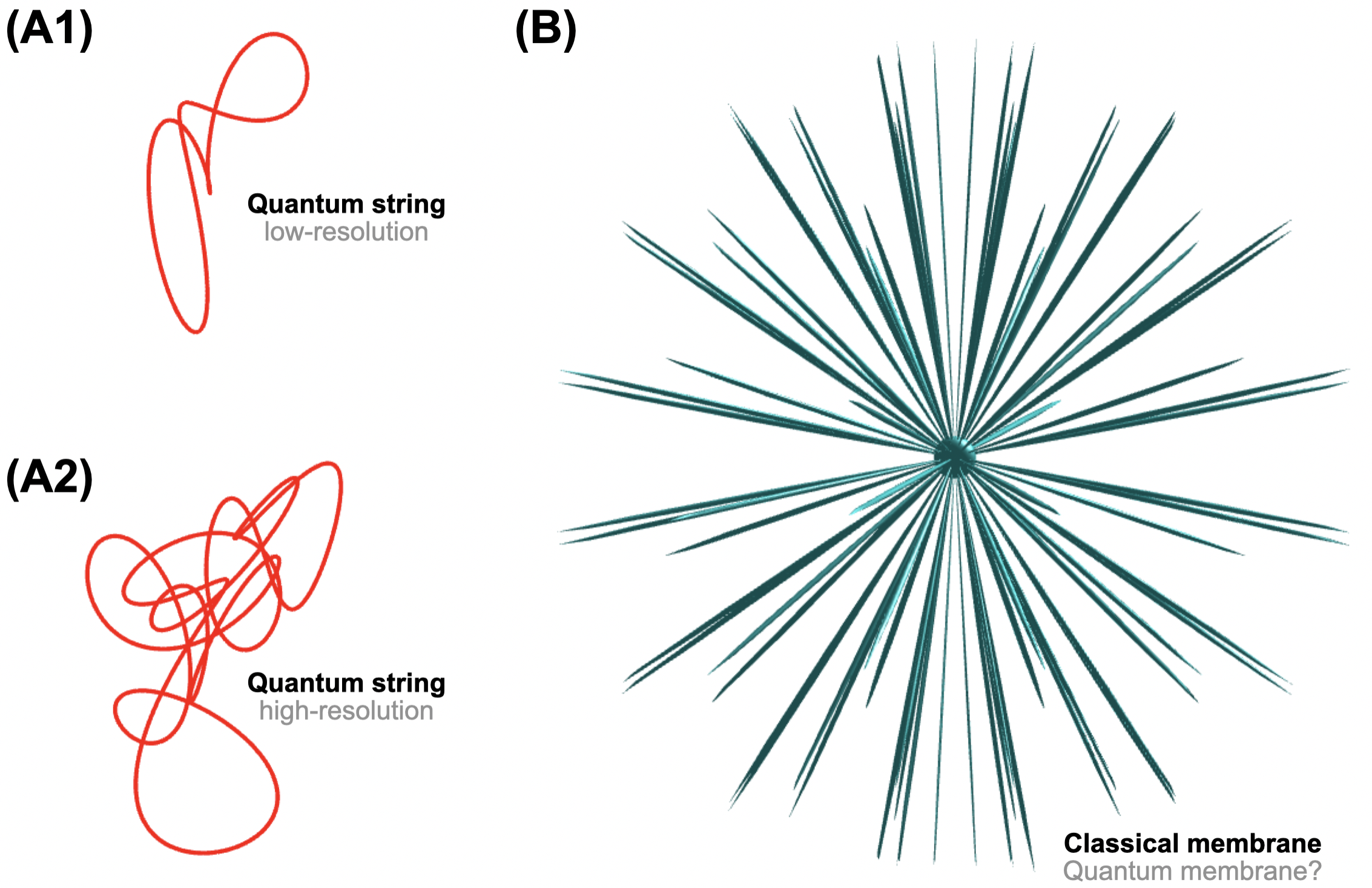}
\caption{\textbf{Visualization of quantum strings and classical membrane.} \textbf{(A1)} A low-resolution snapshot of a quantum string at ground state on a tranverse plane. \textbf{(A2)} A high-resolution snapshot of a quantum string at ground state on a tranverse plane. \textbf{(B)} A spiky shape of a classical membrane to minimize energy. What does a quantum membrane (e.g. at its ground state) look like? The images \textbf{(A1)} and \textbf{(A2)} are inspired by \cite{karliner1988size}, the image \textbf{(B)} is inspired by \cite{taylor2001m,trzetrzelewski2010spiky}.}
\label{fig01}
\end{figure*}

Here we aim to develop visual intuition for quantum membranes. Visual characterization can be crucial for physical insight \cite{gilbert2007visualization}, yet it is often underemphasized in high-energy theory. We analytically and numerically investigate the size and shape statistics of quantum spherical-topology membranes, called \textit{fuzzy spheres} \cite{madore1992fuzzy,arnlind2009fuzzy}, in the $\mathscr{D}=(1+3)$-dimensional spacetime -- the one time and three space dimensions of our observable universe \cite{pardo2018limits}. Historically, such objects were proposed as a lepton model -- specifically, an extensible electron \cite{dirac1962extensible}. Concretely, we investigate the ground state $\Omega$ of regularized membranes described by the SU$(2)$ matrix model \cite{hoppe1980two,hoppe1989quantum}, which is also a reduced Yang-Mills theory \cite{matinyan1981classical,feynman1981qualitative}. We approximate the ground state wavefunction $\Psi_\Omega$ using the Rayleigh-Schr\"{o}dinger variational method \cite{rayleigh1907dynamical,ao2023schrodinger} in an SO$(3)\times$SO$(2)$-invariant space $(U,V)$ that parametrizes membranes \cite{hoppe2023ground}. We show that, at such a low-resolution, when projected onto a two-dimensional transverse plane, the \textit{fuzzy spheres} manifest as ellipses, uniquely defined by their sizes quantified through surface areas $\mathcal{A}$ (or perimeters $\mathcal{L}$) and their asymmetric shapes characterized via eccentricities $\mathcal{E}_3$ (or {\it shape-parameters} $\mathcal{S}$), depending solely on $U$ and $V$. We then estimate the expectation size $\langle \mathcal{A} \rangle_\Omega$, $\langle \mathcal{L} \rangle_\Omega$,  and shape $\langle \mathcal{E}_3 \rangle_\Omega$, $\langle \mathcal{S} \rangle_\Omega$ for the ground state \textit{fuzzy spheres}, representing fundamental geometric constants of quantum membrane mechanics that significantly differ from classical theory (see Fig. \ref{fig01}B). All analytical calculations reported in this work have been checked using \textit{Mathematica 14.3} \cite{Mathematica-14-3}.

\section{The Physics and Mathematics of Fuzzy Spheres}

Consider a $\mathscr{D}$-dimensional spacetime coordinated by $\left( Q_0,Q_1,Q_2,...,Q_{\mathscr{D}-1} \right)$, in which $Q_0$ is the time-like coordinate and $\left(Q_1,Q_2,...,Q_{\mathscr{D}-1} \right)$ are the space-like coordinates. In the light-front frame, the two light-front coordinates are hybrids between the time-like direction and one of the space-like directions, e.g. $Q_\pm = \left( 1/\sqrt{2} \right) \left( Q_0 \pm Q_{\mathscr{D}-1} \right) $. For many relativistic mechanical systems, interpreting one of these coordinates as ``time'' results in the corresponding Hamiltonian being solely dependent on what occurs in the transverse space:
\begin{equation}
\left( Q_1, Q_2, ...,Q_{\mathscr{D}-2}\right) \ .
\label{tranverse_coord}
\end{equation}
Free-moving strings, membranes, and higher-dimensional objects are a few examples \cite{banks1999m,seiberg1997matrix}. In our universe, $\mathscr{D}=4$, so the transverse space is a plane i.e. $\mathscr{D}-2=2$-dimensional. In what follows we relabel the transverse coordinates of Eq. \eqref{tranverse_coord} from $(Q_1,Q_2)$ to $(X,Y)$.

The low-resolution $N=2$ regularized dynamics of relativistic \textit{fuzzy spheres} in $\mathscr{D}=(1+3)$-dimensional spacetime follows a Hamiltonian described by $(\mathscr{D}-2)=2$ traceless Hermitian $N \times N = 2 \times 2$ matrices $\hat{X}$ and $\hat{Y}$, along with their canonical conjugate momentum matrices $\hat{P}_X$ and $\hat{P}_Y$. This Hamiltonian in the center of mass frame, normalized to be half of the relativistically invariant \textit{mass-squared} \cite{hoppe2012relativistic}, is given by:
\begin{equation}
\hat{H} = \frac1{2N} \text{Tr} \left\{ 
 \left( \hat{P}_X^2 + \hat{P}_Y^2 \right) - \left(2\pi N \right)^2 \left[\hat{X},\hat{Y}\right]^2 \right\}  \ ,
\label{Hamiltonian_original}
\end{equation}
where $\text{Tr}\left(\circ \right)$ is the trace operator and $\left[\circ,\circ\right]$ is the commutator. The first term denotes kinetic energy and the second term signifies the interacting potential. The matrices $\hat{X}$, $\hat{Y}$ represent the \textit{fuzzy spheres} fluctuations in the transverse $XY$-plane, and each of them can be mapped independently to $N^2-1=3$ real degrees of freedom denoted as $\vec{x},\vec{y} \in \mathbb{R}^3$:
\begin{equation}
\vec{x}=\left( x_1,x_2,x_3\right) \ , \ \vec{y}=\left( y_1,y_2,y_3\right) \ .
\label{components}
\end{equation}
We provide the details of one such convenient mapping and the consequences followed in Appendix \ref{app:mat_to_comp}. Let us summarize only the relevant results:

\begin{itemize}
    \item \underline{A \textit{fuzzy sphere}} transforms into a two-side ``pancake'' (but no more overlap) when projected onto the transverse $XY$-plane, hence its surface area is twice the area it occupies. If we parametrize this membrane with latitudinal and longitudinal spherical coordinates $(\theta,\phi)$, then points on the \textit{fuzzy sphere} corresponding to the matrices $\hat{X}$, $\hat{Y}$ are located at:
    \begin{equation}
    \begin{split}
        X(\theta,\phi) &= \left( x_1 \cos\phi + x_2 \sin \phi \right) \sin\theta + x_3 \cos\theta \ ,
        \\
        Y(\theta,\phi) &= \left( y_1 \cos\phi + y_2 \sin \phi \right) \sin\theta + y_3 \cos\theta \ ,
        \label{fuzzy_sphere_points}
    \end{split}
    \end{equation}
   identified from the matrix/membrane correspondence -- see Appendix \ref{app:mat_to_comp_MatMem} and note that we have rescaled $\left( \vec{x},\vec{y}\right) \rightarrow \left( 1/\sqrt{3}\right) \left( \vec{x},\vec{y}\right)$.
    \item \underline{The Hamiltonian} in Eq. \eqref{Hamiltonian_original}, for Schr\"{o}dinger wave-function formalism, can be rescaled and written as the following operator:
    \begin{equation}
        \hat{H} = -\frac12 \left( \vec{\nabla}_{\vec{x}}^2 + \vec{\nabla}_{\vec{y}}^2  \right) + \kappa V \ , \ \text{where} \ \kappa = \frac{(4\pi)^2}{\sqrt{3}^6} \ \text{and} \ V \equiv \left( \vec{x} \times \vec{y} \right)^2 \ ,
        \label{Hamiltonian_UV}
    \end{equation}
    in which $\kappa V$ is the interacting potential and $\circ \times \circ$ is the area-product in the $\mathbb{R}^3$ space (see Appendix \ref{app:mat_to_comp_Ham}). There is also a gauge-constraint on the wave-function $\Psi(\vec{x},\vec{y})$:
    \begin{equation}
        \hat{\vec{K}} \Psi = 0 \ , \ \text{where} \ \hat{\vec{K}} = \vec{x} \times \vec{\nabla}_{\vec{x}} + \vec{y} \times \vec{\nabla}_{\vec{y}} \ ,
    \label{gauge_constraint} 
    \end{equation}
    which removes redundant parametrization freedom on the \textit{fuzzy sphere} \cite{hoppe1989quantum,hoppe2012relativistic}.
\end{itemize}

\subsection{The Elliptical Appearance \label{ell_app}}

\begin{figure*}[!htbp]
\includegraphics[width=\textwidth]{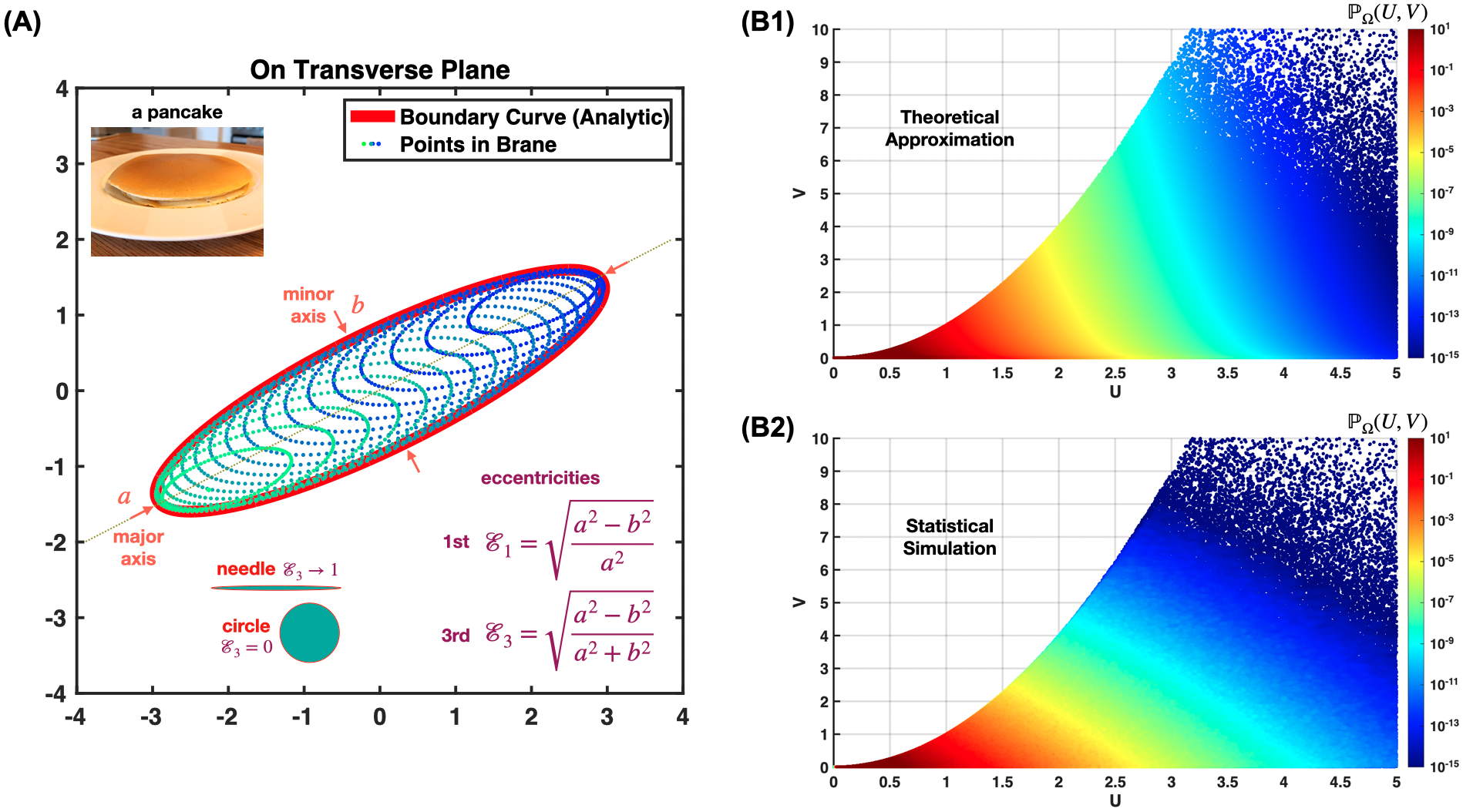}
\caption{\textbf{Elliptical appearance of \textit{fuzzy sphere} and its configuration probability distribution at ground state.} \textbf{(A)} The \textit{fuzzy sphere} appears as an ellipse ``pancake'' in the transverse $XY$-plane, uniquely determined by the size e.g. area $\mathcal{A}$, the shape e.g. eccentricity $\mathcal{E}_3$, modulo a rotation. The scatter points are generated from Eq. \eqref{fuzzy_sphere_points} with the values $(x_1,x_2,x_3,y_1,y_2,y_3)$ drawn at random; each point corresponds to a value of $(\theta,\varphi)$, and the colors range from blue to green as $\theta$ increases (while all $\varphi$ at a given $\theta$ use the same color). \textbf{(B1)} The probability distribution $\mathbb{P}_\Omega (U,V)$, as approximated with Eq. \eqref{probdist} via the Rayleigh-Ritz method. \textbf{(B2)} The probability distribution $\mathbb{P}_\Omega (U,V)$ estimated by a statistical simulation of agent-based random walkers on a fitness landscape (see Appendix \ref{app:gsw}).}
\label{fig02}
\end{figure*}

We define the variables 
\begin{equation}
    U \equiv \frac12 \left( \vec{x}^2 + \vec{y}^2 \right) \ , \ W \equiv VU^{-2} \ ,
\label{define_UW}
\end{equation}
then, from geometric constraints, $W \leq 1$. Also, every points on the $UV$-plane is a SO$(3)\times$SO$(2)$-invariant subspace of the $6$-dimensional $\vec{x}\vec{y}$-space. We derive that $\int dUdV \propto \int d^3\vec{x}d^3\vec{y}$ for invariant integrands in Appendix \ref{app:SO3SO2_int}, which agrees with \cite{hoppe2023ground}. The collection of points described in Eq. \eqref{fuzzy_sphere_points} are bounded by a domain whose boundary in the transverse $XY$-plane is given by the following second-order algebraic equation:
\begin{equation}
   \left( X\vec{y} - Y \vec{x} \right)^2 = V \ ,
\label{2nd_alg}
\end{equation}
which means the \textit{fuzzy sphere} appears as an ellipse -- see Appendix \ref{app:mat_to_ellipse}. It has the surface area $\mathcal{A}$ (two-side) and the eccentricity $\mathcal{E}_3$ equal to:
\begin{equation}
    \mathcal{A}(U,V) = 2\pi V^{1/2} \ , \ \mathcal{E}_3(U,V) = \left( 1-W \right)^{1/4} \ .
\label{AE}
\end{equation}
The perimeter $\mathcal{L}$ and the shape-parameter $\mathcal{S}$ can be estimated (with accuracies better than $4\%$ and $8\%$, respectively) using the following simple formulas:
\begin{equation}
\begin{split}
\mathcal{L}(U,V) \approx \alpha U^{1/2} + \beta V^{1/4} \ \ , \ \ \mathcal{S}(U,V) \equiv \frac{\mathcal{L}^2}{4\pi(\mathcal{A}/2)} \approx \frac{\left[ \alpha U^{1/2} + \beta V^{1/4} \right]}{(2\pi)^2 V^{1/2}} & \ ,
\\
\text{where} \ \ \alpha = 4\sqrt{2} \ \ , \ \ \beta = 2\pi - \alpha &  \ .
\label{LS}
\end{split}
\end{equation}
Note that, instead of the more common 1st eccentricity $\mathcal{E}_1$, here we use the 3rd eccentricity $\mathcal{E}_3=\mathcal{E}_1 \left(2-\mathcal{E}_1^2\right)^{-1/2}$ \cite{collins1980formulas} (also called the normalized polarization \cite{viola2014probability}) to quantify the asymmetry of the ellipse (see Fig. \ref{fig02}A). The \textit{shape-parameter} $\mathcal{S}$ is commonly used in the studies of biological cells and soft matters deformation \cite{boromand2018jamming,wang2021structural,treado2021bridging,bi2016motility}. We present the derivations for Eq. \eqref{AE} and Eq. \eqref{LS} in Appendix \ref{app:mat_to_comp_geo}. When $\vec{x} \perp \vec{y}$, we get $V=U$, $\mathcal{E}_3=0$, and $\mathcal{S}=1$, therefore the \textit{fuzzy spheres} look like circles i.e. perfect-symmetric ellipses. When $\vec{x} \parallel \vec{y}$, we get $V=0$ thus $\mathcal{E}_3=1$ and $\mathcal{S}=\infty$, as the \textit{fuzzy spheres} should have needle-like appearances i.e. maximal-asymmetric ellipses. Note that this is a ``spiky'' shape (to the extent that an ellipse can be), similar to Fig. \ref{fig01}B. Up to a rotation, any pair of elliptical geometric quantities from $\left(\mathcal{A}, \mathcal{L}, \mathcal{E}_3, \mathcal{S}\right)$ besides $\left( \mathcal{E}_3, \mathcal{S}\right)$ can quantify all needs to knows about the \textit{fuzzy sphere} ``pancake''.

\subsection{The Ground State}

From the M-theory supersymmetric matrix model to the simplest SU$(N=2)$ bosonic matrix models, we still know remarkably little about membrane ground states beyond a few special limits -- let alone their excited states \cite{lin2015ground,hoppe1999asymptotic,frohlich2000asymptotic,hoppe2000asymptotic,hoppe2023ground}. However, using the Rayleigh–Ritz method, we can estimate the ground state wavefunction and, within that approximation, even achieve analytical tractability. We choose the normalized test function of the following form:
\begin{equation}
\Psi_{\mu,\nu}(U,V) = \left[4\mu(\mu+\nu)^2\right]^{1/2} \exp \left[ -\left( \mu U + \nu V^{1/2} \right) \right] \ .
\label{RR_groundwf}
\end{equation}
This ansatz -- selected for analytic convenience -- differs from the one used in \cite{hoppe2023ground}. This wavefunction also satisfies the gauge-constraint Eq. \eqref{gauge_constraint} -- see Appendix \ref{app:gauge_constraint}. In other words,
\begin{equation}
\text{$\hat{\vec{K}} \Psi = 0$ selects SO$(3)\times$SO$(2)$-symmetric $\Psi(\vec{x},\vec{y})$.}
\end{equation}
The expectation energy, given Hamiltonian in Eq. \eqref{Hamiltonian_UV}, can be found to be:
\begin{equation}
E_{\mu,\nu} = \int dUdV \left( \Psi_{\mu,\nu}^* \hat{H} \Psi_{\mu,\nu} \right)=\frac12 \left[3\mu + 2\nu + \frac{\nu^2}{\mu} + \frac{3\kappa}{(\mu+\nu)^2} \right] \ ,
\end{equation}
where we have used
\cite{hoppe2023ground}:
\begin{equation}
    \vec{\nabla}_{\vec{x}}^2 + \vec{\nabla}_{\vec{y}}^2 =6\partial_U + 8 U \partial_V +  2 U \partial_U^2 + 8 V \partial_U \partial_V + 8 UV \partial_V^2 \ .
\end{equation}
Minimize this expression $E_{\mu,\nu}$, we obtain the parameters $(\mu_{m},\nu_{m})$:
\begin{equation}
    \mu_{m} = \kappa^{1/3} \times (3/4)^{1/3} \ \ , \ \ \nu_{m} = \left( 2^{1/2}-1\right) \mu_{m} \ ,
\label{minE_param}
\end{equation}
which corresponds to approximating the ground state wavefunction and energy with:
\begin{equation}
\Psi_\Omega(U,V) \approx \Psi_{\mu_{m},\nu_{m}} (U,V) \ , \ E_\Omega \approx E_{\mu_{m},\nu_{m}} = \kappa^{1/3} \times 3(3/4)^{1/3} \ .
\label{estground}
\end{equation}
The probability distribution in the $UV$-space associated with this wavefunction is given by:
\begin{equation}
    \begin{split}
        \mathbb{P}_\Omega(U,V) &= \left| \Psi_\Omega(U,V) \right|^2 =  4\mu_m(\mu_m +\nu_m)^2 \exp \left[ -2\left( \mu_m U + \nu_m V^{1/2} \right) \right] \\
        &= \kappa \times 6\exp\left\{ - 2(3/4)^{1/3} \left[ (\kappa^{1/3} U) + \left(2^{1/2}-1 \right) (\kappa^{2/3} V)^{1/2}\right] \right\} \ ,
\end{split}
\label{probdist}
\end{equation}
as demonstrated in Fig. \ref{fig02}B1.

\section{Expectation Geometry of Quantum Fuzzy Spheres}

\subsection{Theoretical Approximations \label{sec:theo_approx}}

We report some interesting geometric constants of quantum \textit{fuzzy spheres} at ground state where analytical estimations are feasible. We estimate these values by using:
\begin{equation}
    \langle \circ \rangle_\Omega = \int dUdV \mathbb{P}_{\Omega}(U,V) \circ \ .
\label{calculate_expectation}
\end{equation}
For the expectation area $\langle \mathcal{A} \rangle_\Omega$, we use the first expression in Eq. \eqref{AE}:
\begin{equation}
\begin{split}
    \langle \mathcal{A} \rangle_{\Omega} = 2\pi \left(\mu_m+\nu_m \right)^{-1} = \kappa^{-1/3} \times 2\pi (2/9)^{1/6} \approx \kappa^{-1/3} \times 4.890 \ .
\end{split}
\label{A_theo}
\end{equation}
We calculate the expectation eccentricity $\langle \mathcal{E}_3 \rangle_\Omega$, whose equation is given by the second expression of Eq. \eqref{AE}:
\begin{equation}
\begin{split}
    \langle \mathcal{E}_3 \rangle_{\Omega} =& \frac45 \mu_m (\mu_m+\nu_m)^2 \left[-A + \frac{5}{64 \nu_m^{5/2} (\mu_m^2-\nu_m^2)^{7/4}} \left( B_1 + B_2 + B_3 \right) \right] 
    \\
    =& {\scriptstyle \left( \frac{19+13\sqrt{2}}2  \right) -\frac{3\left(\sqrt{2}-1\right)\sqrt{\pi} \Gamma(5/4)}{\Gamma(11/4)} \ _2F_1\left(2,\frac52;\frac{11}4;3-2\sqrt{2}\right) + \frac3{8} \left( \frac{299249 + 211601 \sqrt{2}}{2}  \right)^{1/4} \times }
    \\
    & {\scriptstyle  \left( \frac12 \ln \left\{ \frac{2-2\left[ 2 (-1+\sqrt{2})\right]^{1/4} + \left[ 2 (-1+\sqrt{2})\right]^{1/2}}{2+2\left[ 2 (-1+\sqrt{2})\right]^{1/4} + \left[ 2 (-1+\sqrt{2})\right]^{1/2}} \right\} + \left\{\text{arccot}\left[ 1-2^{3/4} (1+\sqrt{2})^{1/4}\right]-\text{arccot}\left[ 1+2^{3/4} (1+\sqrt{2})^{1/4}\right] \right\} \right) }
    \\
    \approx & 8.337 \times 10^{-1} \ ,
\end{split}
\label{E_theo}
\end{equation}
where the coefficients $A$ and $B_1$, $B_2$, $B_3$ are:
\begin{equation}
\begin{split}
    A &= \frac{3\sqrt{\pi} \nu_m \Gamma(9/4) \ _2F_1 \left( 2, 5/2;11/4;\nu_m^2/\mu_m^2\right)}{2\mu_m^4 \Gamma(11/4)} \ , 
    \\ 
    B_1 &= 8\mu^{-1}_m \nu_m^{1/2} \left(3\mu_m^2 - 2 \nu_m^2 \right) \left( \mu_m^2 - \nu_m^2 \right)^{3/4} \ ,
    \\
    B_2 &= 6\sqrt{2} \mu_m \left( \mu_m^2 - 2\nu_m^2 \right) {\scriptstyle \left\{ \text{arctan} \left[ 1- \left(\frac{2\mu_m}{\nu_m}\right)^{1/2} \left( 1-\frac{\nu_m^2}{\mu_m^2} \right)^{1/4} \right] - \text{arccot}\left[ 1+ \left(\frac{2\mu_m}{\nu_m}\right)^{1/2} \left( 1-\frac{\nu_m^2}{\mu_m^2} \right)^{1/4} \right] \right\}} \ , 
    \\
    B_3 &= 3\sqrt{2} \mu_m^3 \left\{1 + \frac{2\nu_m^2}{\mu_m^2} \ln \left[ \frac{\nu_m - (2\nu_m)^{1/2} (\mu_m^2-\nu_m^2)^{1/4} + (\mu_m^2-\nu_m^2)^{1/2}}{\nu_m + (2\nu_m)^{1/2} (\mu_m^2-\nu_m^2)^{1/4} + (\mu_m^2-\nu_m^2)^{1/2}} \right] \right\} \ .
\end{split}
\end{equation}
Note that, $\Gamma(...)$ is the Gamma-function and $_2F_1(...)$ is the hypergeometric function.

\begin{figure*}[!htbp]
\includegraphics[width=\textwidth]{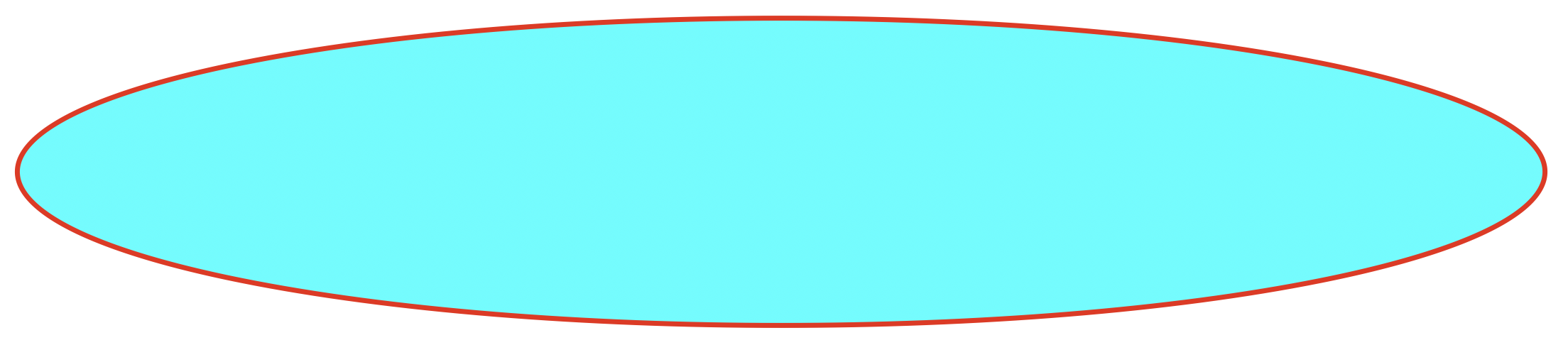}
\caption{\textbf{The typical shape of ground state quantum \textit{fuzzy spheres} at low-resolution.} The aspect ratio (the ratio between the lengths of the major and minor axes) of this ``pancake'' is about $4.973$.}
\label{fig03}
\end{figure*}

For the expectation perimeter $\langle \mathcal{L} \rangle_\Omega$, we utilize the first expression in Eq. \eqref{LS}, which serves as a high-accuracy approximation rather than an exact formula:
\begin{equation}
\begin{split}
    \langle \mathcal{L} \rangle_{\Omega} &\approx \frac{1}{4} \left( \frac{\pi}2 \right)^{1/2} \left\{ \frac{-2\alpha \mu_m^{5/2} - 5\alpha \mu_m^{3/2} \nu_m + 3\beta \mu_m^{1/2} \nu_m^2 + 2\alpha \left( \mu_m+\nu_m \right)^{5/2}}{\nu_m^2 \left[ \mu_m \left( \mu_m+\nu_m \right) \right]^{1/2}} \right\} \\
    & = \kappa^{-1/6} \times {\scriptstyle \frac14 \left( \frac{10-7\sqrt{2}}{3}\right)^{1/6} \left[ \left(41+29\sqrt{2} \right)\pi \right]^{1/2} \left[ 4\left(-6+\sqrt{2}+4\sqrt[4]{2}\right) + 3 \left( -4+3\sqrt{2} \right) \pi  \right] }
    \\
    & \approx \kappa^{-1/6} \times 6.789 \ .
\end{split}
\label{L_theo}
\end{equation}
Finally, to obtain the expectation \textit{shape parameter} $\langle \mathcal{S} \rangle_\Omega$, we need the second expression in Eq. \eqref{LS}, which is also an approximation:
\begin{equation}
\begin{split}
    \langle \mathcal{S} \rangle_{\Omega}
    & \approx \frac1{4\pi^2}\left\{ \frac{ \splitfrac{\left[ \alpha^2 \nu_m \left(2\mu_m+\nu_m \right) + \alpha \beta \mu_m \left( -\mu_m + \nu_m \right) + \beta^2 \mu_m \nu_m \right] \nu_m^{1/2} \ \ \ \ \ \ \ \ \ \ \ \ }{+ \alpha\beta \mu^{1/2} \left( \mu_m + \nu_m \right)^{2} \arctan \left( \sqrt{\nu_m/\mu_m} \right)}}{\mu_m \nu_m^{3/2}} \right\} 
    \\
    &= {\scriptstyle \frac1{\pi^2} \left\{ \left(16 + 16\sqrt{2} - 4\pi - 4\sqrt{2} \pi + \pi^2\right) + 2
    \left( 7+5\sqrt{2}\right)^{1/2} \left( -4+\sqrt{2}\pi\right) \text{arctan}\left[2\left(1+\sqrt{2}\right) \right]^{1/2} \right\} }
    \approx 2.225  \ .
    \\
\end{split}
\label{S_theo}
\end{equation}
In Fig. \ref{fig03} we show an ellipse with the \textit{shape parameter} equal to $\langle \mathcal{S}\rangle_\Omega$. This is the most typical shape of ground state quantum \textit{fuzzy spheres}, which is not so circular but far from being spiky (classical expectation).

It is also informative to examine how these geometric expectation values for the ground state fuzzy sphere differ from those obtained when the parameters $(\vec{x},\vec{y})$ are selected randomly from a six-dimensional Gaussian distribution. This baseline helps identifying which geometric signatures are intrinsic to the ground state, as opposed to generic statistical fluctuations. We study this in Appendix \ref{app:Gauss}.

\subsection{Statistical Experimental Results  \label{sec:stat_sim}}

\begin{figure*}[!htbp]
\includegraphics[width=\textwidth]{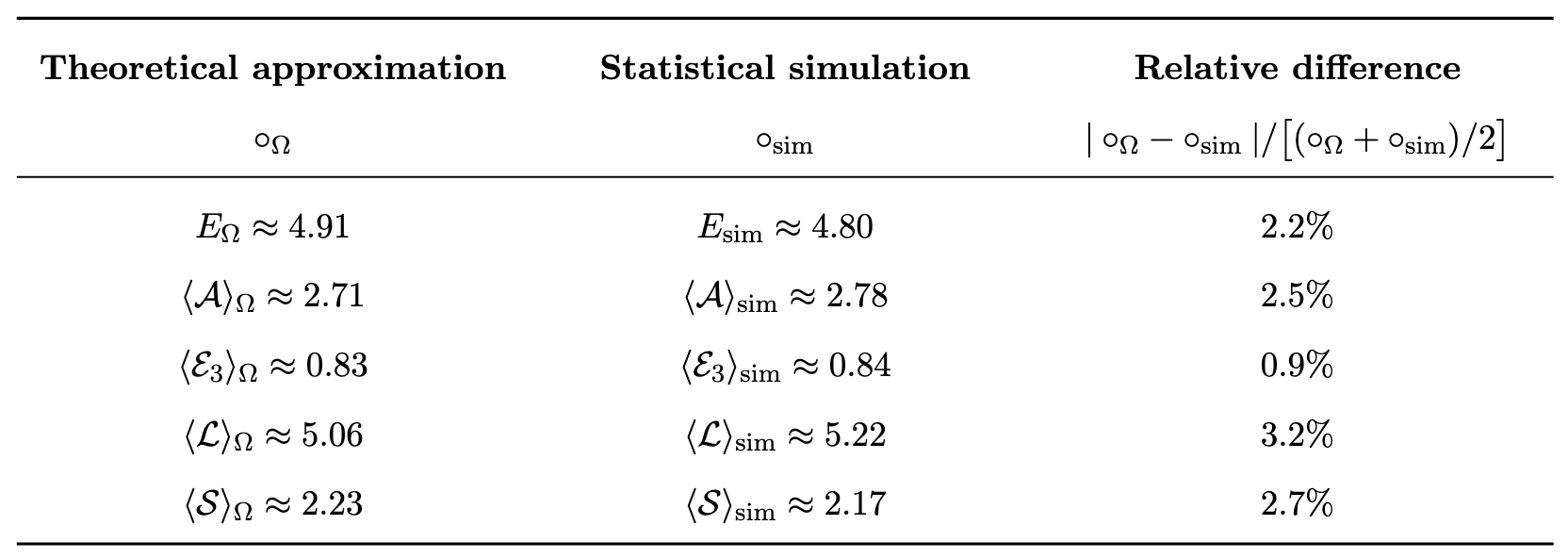}
\caption{\textbf{Comparison between the estimations from theoretical approximation and from the statistical simulation.} We report the numerical expectation values for the fuzzy sphere area $\mathcal{A}$, the eccentricity $\mathcal{E}_3$, the perimeter $\mathcal{L}$, and the shape parameter $\mathcal{S}$ -- see Eq. \eqref{AE} and Eq. \eqref{LS}. We also report the numerical estimation for the ground state energy $E$.}
\label{figT01}
\end{figure*}

The ground state can also be found from doing an agent-based simulation of random walkers on a fitness landscape \cite{ao2023schrodinger}. It should be noted that there are many approaches to numerically estimate the ground state of Eq. \eqref{Hamiltonian_UV}, but this approach is especially is especially simple, robust, and stable to perform this task. Agent-based simulations, in general, offer \textit{adaptive resolution}, where dense regions yield more samples while sparse regions take in information from larger neighborhoods. They also avoid common grid and mesh artifacts (e.g., numerical diffusion and the Courant–Friedrichs–Lewy constraint) which can lead to instability \cite{leveque2002finite}, and they stay symmetric by construction (e.g. no grid-alignment bias). They scale better at high-dimensional space -- six dimensions, as in Eq. \eqref{Hamiltonian_UV}, is high -- without the exponential growth of degrees of freedom and meshing complexity that burdens grid- and mesh-based methods. We explain the details of our simulation in Appendix \ref{app:gsw}.

We show the estimated ground state probability distribution $\mathbb{P}_\Omega(U,V)$ obtained from the simulation in Fig. \ref{fig02}B2. Qualitatively, it closely resembles the theoretically approximated $\mathbb{P}_\Omega(U,V)$, with the most noticeable discrepancy due to functional behavior along the boundary $V=U^2$ -- it seems to suggest a Neumann (zero-gradient) boundary condition there. Quantitatively, we compare the resulting estimates for the expectation values of geometric quantities introduced in Section \ref{ell_app} -- computed using this empirical distribution -- with the theoretical approximations given in Eqs. \eqref{A_theo}, \eqref{E_theo}, \eqref{L_theo}, and \eqref{S_theo} in Fig. \ref{figT01}. All deviations are below $3.2\%$, demonstrating the high accuracy of our theoretical approximations.

\section{Discussion}

At low-resolution, quantum \textit{fuzzy spheres} in our $\mathscr{D}=(1+3)$-dimensional universe can be described by the regularized SU$(N=2)$ matrix model. 
In the transverse plane, they exhibit a simple geometric structure, which takes the form of ellipses. By approximating its ground state with the Rayleigh-Ritz method, we can fully determine the expectation geometry, including both size and shape. We hope this study is useful and timely, given that recent developments in both experimental and theoretical physics have revitalized matrix models as a \textit{lingua franca} across multiple subfields, notably in analog simulation of quantum gravity \cite{maldacena2023simple} and quantum cognitive machine learning \cite{candelori2025robust}. Future works must extend beyond the ground state, consider different membrane topology, and incorporate supersymmetry to obtain findings more relevant to high energy physics.

\section{Acknowledgement}

We thank Ramzi R. Khuri for the encouragement to report our findings to a wider audience. We also thank Huy D. Tran, Van H. Do, Long T. Nguyen, and Shi Chen for their suggestions during the beginning stage of this work. 

\appendix

\section{Components of Matrix \label{app:mat_to_comp}}

The components in Eq. \eqref{components} are those of $\hat{X}$, $\hat{Y}$ in the basis of $N=2$-dimensional representation of the spherical rotation group SO$(3)$ i.e. $N^2-1=3$ matrices $\hat{T}_1$, $\hat{T}_2$, $\hat{T}_3$ with the normalization $\text{Tr}\left( \hat{T}_a \hat{T}_b \right) = N \delta_{ab} = 2\delta_{ab}$. A simple choice is the familiar set of Pauli matrices $\hat{\sigma}_1$, $\hat{\sigma}_2$, $\hat{\sigma}_3$, which can be found in many quantum mechanics textbooks \cite{sakurai1995modern,landau2013quantum}:
\begin{equation}
\begin{split}
\hat{X} = x_1 \hat{\sigma}_1 + x_2 \hat{\sigma}_2 + x_3 \hat{\sigma}_3 = \vec{x} \cdot \hat{\vec{\sigma}} \ , \ \hat{Y} = y_1 \hat{\sigma}_1 + y_2 \hat{\sigma}_2 + y_3 \hat{\sigma}_3 = \vec{y} \cdot \hat{\vec{\sigma}} \ .
\end{split}
\label{get_matrix}
\end{equation}
The same goes for $\left( \vec{p}_{\vec{x}},\vec{p}_{\vec{y}}\right)$ as components of $\left(\hat{P}_X, \hat{P}_Y \right)$. 

\subsection{The Matrix/Membrane Correspondence \label{app:mat_to_comp_MatMem}}

Using the matrix/membrane correspondence, in which each spherical harmonic function \cite{blanco1997evaluation} defined on the angular S$^2$ parametrization is mapped to a $SO(3)$ generator i.e. $\sqrt{4\pi} \mathbb{Y}_{1,m}(\theta,\phi) \leftrightarrow \hat{\sigma}_a$ \cite{hoppe2012relativistic}, we can map functions describing all positions on the membrane in the transverse plane coordinates $\Big(X(\theta,\phi),Y(\theta,\phi) \Big)$ to two matrices $\left( \hat{X},\hat{Y} \right)$. The Hamiltonian in the functional form derived from the action of the \textit{fuzzy sphere} in spacetime \cite{hoppe1989quantum} can become the matrix form in Eq. \eqref{Hamiltonian_original} by considering the following mapping:
\begin{equation}
\begin{split}
    a=1 \leftrightarrow \ m=1 \ &\text{:} \ \sqrt{4\pi} \mathbb{Y}_{1,1} = \sqrt{3}\cos\phi \sin\theta \ , 
    \\
    a=2 \leftrightarrow \ m=-1 \ &\text{:} \ \sqrt{4\pi} \mathbb{Y}_{1,-1} = \sqrt{3}\sin\phi \sin\theta \ , 
    \\
    a=3 \leftrightarrow \ m=0 \ &\text{:} \ \sqrt{4\pi} \mathbb{Y}_{1,0} = \sqrt{3} \cos\theta \ .
\end{split}
\end{equation}
Plug these into Eq. \eqref{get_matrix},
we can recover Eq. \eqref{fuzzy_sphere_points} after rescaling $\left( \vec{x},\vec{y}\right) \rightarrow \left( 1/\sqrt{3}\right) \left( \vec{x},\vec{y}\right)$.

\subsection{The Hamiltonian \label{app:mat_to_comp_Ham}}

From the commutator/are-product relationship for Pauli vectors \cite{sakurai1995modern,landau2013quantum}:
\begin{equation}
    \left[ \hat{X},\hat{Y} \right] = \left[ \vec{x}\cdot \hat{\vec{\sigma}},\vec{y}\cdot \hat{\vec{\sigma}}\right] = 2i \left(\vec{x} \times \vec{y} \right) \cdot \hat{\vec{\sigma}} \ ,
\end{equation}
we plug it into Eq. \eqref{Hamiltonian_original} to arrive at:
\begin{equation}
        \hat{H} = \frac12 \left( \vec{p}_{\vec{x}}^2 + \vec{p}_{\vec{y}}^2  \right) + \left( 4\pi \right)^2 \left( \vec{x} \times \vec{y} \right)^2 \ .
\end{equation}
Promote the canonical conjugate momentum $\left(\vec{p}_{\vec{x}},\vec{p}_{\vec{y}}\right)$ of the degrees of freedom $\left( \vec{x},\vec{y} \right)$ into gradient operators $-i \left(\vec{\nabla}_{\vec{x}}, \vec{\nabla}_{\vec{y}}\right)$, we get Eq. \eqref{Hamiltonian_UV} after rescaling $\left( \vec{x},\vec{y}\right) \rightarrow \left( 1/\sqrt{3}\right) \left( \vec{x},\vec{y}\right)$ (as already done in Appendix \ref{app:mat_to_comp_MatMem}) and $\hat{H} \rightarrow 3\hat{H}$.

\section{From SO$(3)\times$SO$(2)$ Parametrization to Elliptical Geometric Quantities}

\subsection{Integral Measure \label{app:SO3SO2_int}}

Eq. \eqref{components} is equivalent to:
\begin{equation}
    \vec{x} = x_1 \vec{e}_1 + x_2 \vec{e}_2 + x_3 \vec{e}_3 \ , \ \vec{y} = y_1 \vec{e}_1 + y_2 \vec{e}_2 + y_3 \vec{e}_3 \ ,
\end{equation}
where the set of three unit-vectors $\{\vec{e}_1,\vec{e}_2,\vec{e}_3 \}$ is the standard Cartesian orthogonal basis of $\mathbb{R}^3$ space. Let us define a spherical coordinate system $(x,\theta_x,\varphi_x)$ to describe $\vec{x}$, where $x=|\vec{x}|$, $\theta_x$ is the angle between $\vec{x}$ and $\vec{e}_3$, and $\varphi_x$ is the azimuthal angle measured in the $\vec{e}_2\vec{e}_3$-plane from $\vec{e}_1$. For $\vec{y}$, for convenience in later calculations we introduce another spherical coordinate system $(y,\theta_y,\varphi_y)$, in which $y=|\vec{y}|$, $\theta_y$ is the angle between $\vec{y}$ and $\vec{x}$, and $\varphi_y$ is the azimuthal angle measured in the $\vec{e}_2\vec{n}$-plane from $\vec{n}$ (with $\vec{n}=\vec{e}_2 \times \vec{x}$). Then, the mapping between the coordinate values $(x_1,x_2,x_3,y_1,y_2,y_3)$ and $(x,\theta_x,\varphi_x,y,\theta_y,\varphi_y)$ is:
\begin{equation}
\begin{split}
    x_1 &= x \sin \theta_x \cos\varphi_x \
     \ , \ x_2 = x\sin\theta_x \sin\varphi_x \ , 
    \ x_3 = x \cos\theta_x \ , 
    \\
    y_1 &= y(\sin\theta_x \cos\theta_y \cos\varphi_x + \cos\theta_x \sin\theta_y \cos\varphi_x \cos\varphi_y - \sin\theta_y \sin\varphi_x \sin\varphi_y) \ ,
    \\
    y_2 &=y(\sin\theta_x \cos\theta_y \sin\varphi_x + \cos\theta_x \sin\theta_y \sin\varphi_x \cos\varphi_y + \sin\theta_y \cos\varphi_x \sin\varphi_y) \ ,
    \\
    y_3 &=y(\cos\theta_x \cos\theta_y - \sin\theta_x \sin\theta_y \cos\varphi_y) \ .
\end{split}
\label{xy_Cart_to_Sphere}
\end{equation}

Consider the integral measure
$$\int d\omega = \int dx_1 dx_2 dx_3 dy_1 dy_2 dy_3 \ , $$
which, after the change of variables, becomes:
\begin{equation}
    \int d\omega = \int \det \left( J \right) dx d\theta_x d\varphi_y dy d\theta_y d\varphi_y \ .
\label{int_measure_in_xy_sphere}
\end{equation}
The matrix $J$ is the $6\times 6$ Jacobian, which can be rewritten in $4$ blocks of $3 \times 3$ as follows:
\begin{equation}
    J =  \frac{\partial (x_1,x_2,x_3,y_1,y_2,y_3)}{\partial (x,\theta_x,\varphi_x,y,\theta_y,\varphi_y)} = \begin{pmatrix} \frac{\partial (x_1,x_2,x_3)}{\partial (x,\theta_x,\varphi_x)} & \frac{\partial (x_1,x_2,x_3)}{\partial (y,\theta_y,\varphi_y)} \\ \frac{\partial (y_1,y_2,y_3)}{\partial (x,\theta_x,\varphi_x)} & \frac{\partial (y_1,y_2,y_3)}{\partial (y,\theta_y,\varphi_y)} \end{pmatrix} \ .
\end{equation}
Since $\partial (y_1,y_2,y_3)/\partial (x,\theta_x,\varphi_x)=0$, the Jacobian $J$ is a block-triangular, therefore:
\begin{equation}
    \det (J) = \det \left( \frac{\partial (x_1,x_2,x_3)}{\partial (x,\theta_x,\varphi_x)} \right) \det \left( \frac{\partial (y_1,y_2,y_3)}{\partial (y,\theta_y,\varphi_y)} \right) \ .
\end{equation}
Both of these factors are Jacobians for change of variables from Cartesian to spherical coordinates, which give:
\begin{equation}
\begin{split}
    \det \left( \frac{\partial (x_1,x_2,x_3)}{\partial (x,\theta_x,\varphi_x)} \right) = x^2 \sin\theta_x \ , \  \det \left( \frac{\partial (y_1,y_2,y_3)}{\partial (y,\theta_y,\varphi_y)} \right) = y^2\sin\theta_y &
    \\
    \Longrightarrow \ \ \det(J) = x^2 y^2 \sin \theta_x \sin \theta_y  & \ .
    \end{split}
\label{int_measure_length_angle}
\end{equation}
We then use this result in Eq. \eqref{int_measure_in_xy_sphere} and integrate out $(\theta_x,\varphi_x,\varphi_y)$ to obtain:
\begin{equation}
\begin{split}
    \int d\omega &= \int x^2 y^2 \sin\theta_y dx dy d\theta_y \int_0^{\pi} \sin\theta_x d\theta_x \int_0^{2\pi} d\varphi_x \int^{2\pi}_0 d\varphi_y 
    \\
    &= 8\pi^2 \int x^2 y^2 \sin\theta_y dx dy d\theta_y = 2\pi^2 \int \frac{xy}{(1-\sin^2\theta_y)^{1/2}} d(x^2) d(y^2) d(\sin^2 \theta_y) \ .
\end{split}
\end{equation}
Note that the scan of $\theta_y$ from $0$ to $2\pi$ doubly-cover the scan of $\sin^2\theta_y$ from $0$ to $1$. 

The variables $V$ and $U$ defined in Eq. \eqref{Hamiltonian_UV} and Eq. \eqref{define_UW} can be expressed as:
\begin{equation}
    U = \frac12 (x^2 + y^2) \  , \ V = x^2 y^2 \sin\theta_y^2 \ .
\end{equation}
Define a variable $T=x^2$, then $y^2=2U-T$ and $\sin^2\theta_y = V/T(2U-T)$. Following Eq. \eqref{int_measure_length_angle} we can rewrite the integral measure as:
\begin{equation}
    \int d\omega = 2\pi^2 \int \left( \frac{T(2U-T)}{1-\frac{V}{T(2U-T)}} \right)^{1/2} \det (\tilde{J})dT dU dV \ ,
\label{int_measure_TUV}
\end{equation}
where the Jacobian $\tilde{J}$ is given by a $3 \times 3$ matrix $\partial(x^2,y^2,\sin^2\theta_y) / \partial(T,U,V)$. The determinant of this Jacobian can be calculated to be:
\begin{equation}
    \det (\tilde{J}) = \frac{8}{T(2U-T)} \ , 
\end{equation}
hence Eq. \eqref{int_measure_TUV} becomes:
\begin{equation}
    \int d\omega = 16\pi^2 \int \left[ (U^2-V) - (U-T)^2 \right]^{-1/2} dT dU dV \ .
\end{equation}
After integrating out the variable $T$, we arrive at:
\begin{equation}
\begin{split}
    \int d\omega &= 16\pi^2 \int dU dV \int_{(U-T)^2 \leq U^2-V} \left[ (U^2-V) - (U-T)^2 \right]^{-1/2} dT
    \\
    &= 16\pi^2 \int dU dV \int^{+(U^2-V)^{1/2}}_{-(U^2-V)^{1/2}} \left[ (U^2-V) - (U-T)^2 \right]^{-1/2} d(U-T)\ .
    \\
    &= 16\pi^2 \int dU dV \int^{+\pi/2}_{-\pi/2} d\Theta \Big|_{(U-T)=(U^2-V)^{1/2} \sin\Theta}  = 16\pi^3 \int dU dV \ .
\end{split}
\end{equation}
Thus, we have shown that $\int d^3 \vec{x} d^3\vec{y}= 16\pi^3 \int dUdV$. It is important to note that, when we do the integral over the whole $UV$-space, we mean to calculate the following operation:
\begin{equation}
    \int dU dV \equiv \int_0^{\infty} dU \int_0^{U^2} dV \ .
\end{equation}

\subsection{An Elliptical Envelope \label{app:mat_to_ellipse}}

We start from the Eq. \eqref{fuzzy_sphere_points} that maps a point on the fuzzy sphere at the geographic coordinate $(\theta,\varphi)$ to a physical position $(X,Y)$ on the transverse plane. These equations can be rewritten as:
\begin{equation}
    \left( \frac{X-x_3 \cos\theta}{\sin\theta} \right) = x_1 \cos\varphi + x_2 \sin\varphi
\label{eq1}
\end{equation}
and
\begin{equation}
\left( \frac{Y-y_3 \cos\theta}{\sin\theta} \right) = y_1 \cos\varphi + y_2 \sin\varphi \ .
\label{eq2}
\end{equation}
We multiply both sides of Eq. \eqref{eq1} with $y_1$ and both sides of Eq. \eqref{eq2} with $x_1$, then subtract them to arrive at:
\begin{equation}
    \sin\varphi = \frac{y_1 \left( \frac{X-x_3\cos\theta}{\sin\theta}\right) - x_1 \left( \frac{Y-y_3\cos\theta}{\sin\theta} \right)}{y_1 x_2 - x_1 y_2} \ . 
\label{eq3}
\end{equation}
Similarly, we multiply both sides of Eq. \eqref{eq1} with $y_2$ and both sides of Eq. \eqref{eq2} with $x_2$, then subtract them to obtain:
\begin{equation}
    \cos\varphi = \frac{y_2 \left( \frac{X-x_3\cos\theta}{\sin\theta}\right) - x_2 \left( \frac{Y-y_3\cos\theta}{\sin\theta} \right)}{y_2 x_1 - x_2 y_1} \ . 
\label{eq4}
\end{equation}
Since $\sin^2\varphi + \cos^2\varphi = 1$, Eq. \eqref{eq3} and Eq. \eqref{eq4} together give us an implicit function $F_\theta (X,Y)=0$ that describes the set of points with $\varphi \in [0,2\pi]$ in the transverse $XY$-space for a specific value of $\theta$, i.e.
\begin{equation}
    F_\theta(X,Y) = \frac{\alpha - \beta t + \gamma t^2 }{1-t^2} -1 = 0 \ ,
\label{implicit_curve}
\end{equation}
where $t=\cos\theta$ and 
\begin{equation}
\begin{split}
    \alpha &= \left( \frac{y_1 X - x_1 Y}{y_1 x_2 - x_1 y_2} \right)^2 + \left( \frac{y_2 X - x_2 Y}{y_2 x_1 - x_2 y_1} \right)^2 \ ,
    \\
    \beta &= 2 \left[ \frac{(y_1 X - x_1 Y)(y_1 x_3 - x_1 y_3)}{(y_1 x_2 - x_1 y_2)^2} + \frac{(y_2 X - x_2 Y)(y_2 x_3 - x_2 y_3)}{(y_2 x_1 - x_2 y_1)^2} \right] \ ,
    \\
    \gamma &=\left( \frac{y_1 x_3 - x_1 y_3}{y_1 x_2 - x_1 y_2} \right)^2 + \left( \frac{y_2 x_3 - x_2 y_3}{y_2 x_1 - x_2 y_1} \right)^2 \ .
\end{split}
\label{abc}
\end{equation}
The envelope of all these $F_\theta (X,Y)=0$ curves is the outer boundary of the fuzzy sphere projection in the transverse $XY$-space. To find the equation that describes this envelope, we solve for the set of points $(X,Y)$ that simultaneously satisfy $F_\theta(X,Y)=0$ and $\partial_\theta F_\theta(X,Y)=0$ \cite{bruce1984curves}. Since $\partial_\theta F_\theta(X,Y)=0$ also means $\partial_t F_\theta(X,Y)=0$, we can use it to represent $t$ as the function of $A$, $B$ (which are functions of $X$ and $Y$), and $C$ (which is a constant for a given $\vec{x}$ and $\vec{y}$):
\begin{equation}
\begin{split}
    \partial_t F_\theta(X,Y) = \frac{2(\alpha+\gamma)t - \beta (1+t^2)}{(1-t^2)^2}= 0 &
    \\ 
    \Longrightarrow \ \ t = \frac{(\alpha+\gamma) \pm \left[ (\alpha+\gamma)^2 - \beta^2\right]^{1/2}}{\beta} \ . 
\end{split}
\end{equation}
Take this value of $t$ and plug it in Eq. \eqref{implicit_curve}, we get the following:
\begin{equation}
\begin{split}
    F_{\theta}(X,Y) \Big|_{t : \ \partial_t F_{\theta}(X,Y) = 0} = \frac{(\alpha-\gamma) \pm \left[ (\alpha +\gamma)^2 - \beta^2 \right]^{1/2}}{2} -1 = 0 &
    \\
    \Longrightarrow \ \ \alpha - \frac{\beta^2}{4(1+\gamma)} = 1 & \ .
\end{split}
\end{equation}
From the definitions of $\alpha$, $\beta$, and $\gamma$ in Eq. \eqref{abc}, after some algebraic manipulation (i.e. clearing denominators), we can rewrite the above finding as:
\begin{equation}
V(w_1^2 +w_2^2) - (v_{13} w_1 + v_{23} w_2)^2 = V v_{12}^2 \ ,
\label{envelope_curve}
\end{equation}
where $V$ is as given in Eq. \eqref{Hamiltonian_UV},
\begin{equation}
    v_{jk} = y_j x_k - x_j y_k \ \ \text{so that} \ \ V^2 = v_{12}^2 + v_{23}^2 + v_{31}^2 \ \ \text{, and} \ \  w_j = y_j X - x_j Y  \ ,
\end{equation}
with $j \in \{ 1,2,3\}$. We can further rewrite Eq. \eqref{envelope_curve} as:
\begin{equation}
\begin{split}
    V v_{12}^2 &= (v_{12}^2 + v_{23}^2) w_1^2 + (v_{12}^2 + v_{13}^2) w_2^2 - 2 v_{13} v_{23} w_1 w_2
    \\
    &=\left[ y_1^2 (v_{12}^2 + v_{23}^2) + y_2^2 (v_{12}^2 + v_{13}^2) - 2y_1 y_2 v_{13} v_{23} \right]X^2 
    \\
    & \ \ +\left[ x_1^2 (v_{12}^2 + v_{23}^2) + x_2^2 (v_{12}^2 + v_{13}^2) - 2x_1 x_2 v_{13} v_{23}  \right]Y^2
    \\
    &\ \ -2 \left[ x_1 y_1 (v_{12}^2 + v_{23}^2) + x_2 y_2 (v_{12}^2 + v_{13}^2) - (x_1 y_2 + x_2 y_1) v_{13} v_{23} \right] XY \ .
\end{split}
\end{equation}
The coefficient before $X^2$ can be simplified into $\vec{y}^2 v_{12}^2$, the coefficient before $Y^2$ can be simplified into $\vec{x}^2 v_{12}^2$, and coefficient before $XY$ can be simplified into $-2(\vec{x}.\vec{y})v_{12}$, and thus:
\begin{equation}
\begin{split}
    V v_{12}^2 = \vec{y}^2 v_{12}^2 X^2 + \vec{x}^2 v_{12}^2 Y^2 -2(\vec{x}.\vec{y})v_{12} XY &
    \ \ \Longrightarrow \ \ V = (\vec{y} X - \vec{x} Y )^2 \ .
\end{split}
\end{equation}
This is identical with Eq. \eqref{2nd_alg}. The envelope is a conic curve. To be more precise, a general conic curve can be described by a quadratic equation:
\begin{equation}
    A X^2 + B XY + C Y^2 + DX + EY + F = 0 \ , 
\label{conic_curve}
\end{equation}
and our envelope has $A = \vec{y}^2$, $B=-2(\vec{x}.\vec{y})$, $C = \vec{x}^2$, $D=E=0$, and $F=V$. Since our envelope always satisfies $4AC\geq B^2$, it is an ellipse. 

\subsection{Geometric Interpretation \label{app:mat_to_comp_geo}}

The aim of this section is to derive the formula for the area $\mathcal{A}$, the third-eccentricity $\mathcal{E}_3$, and the approximate formula for the perimeter $\mathcal{L}$ as functions of variables $U$ and $V$. Note that for the shape parameter $\mathcal{S}$, that value can be calculated directly from $\mathcal{A}$ and $\mathcal{L}$.

For an ellipse described by Eq. \eqref{conic_curve} in which $D=E=0$, its area is given by \cite{wrede2002theory}:
\begin{equation}
    \mathcal{A}_1 = \frac{2\pi F}{(4 AC - B^2)^{1/2}} \ .
\end{equation}
For $A = \vec{y}^2$, $B=-2(\vec{x}.\vec{y})$, $C = \vec{x}^2$, and $F=V$, we arrive at:
\begin{equation}
    \mathcal{A}_1 = \frac{\pi V}{\left[ \vec{x}^2 \vec{y}^2 - (\vec{x}.\vec{y})^2 \right]^{1/2}} \ .
\end{equation}
But note that $\vec{x}^2 \vec{y}^2 = x^2 y^2$, $(\vec{x}.\vec{y})^2 = xy \cos\theta_y$ (the value $x$, $y$, and $\theta_y$ is as defined in Appendix \ref{app:SO3SO2_int}), therefore $\vec{x}^2 \vec{y}^2 - (\vec{x}.\vec{y})^2 = x^2 y^2 \sin^2\theta_y$, which is exactly equal to $(\vec{x} \times \vec{y})^2 = V$, i.e.
\begin{equation}
   \vec{x}^2 \vec{y}^2 - (\vec{x}.\vec{y})^2 = V
\label{also_V}
\end{equation}
Thus, $\mathcal{A}_1 = \pi V^{1/2}$, and since the projected area of the fuzzy sphere has twice as much the area of this ellipse, we obtain $\mathcal{A} = 2\mathcal{A}_1 = 2 \pi V^{1/2} $.

The formula for the first-eccentricity of an ellipse described by Eq. \eqref{conic_curve} when $D=E=0$ is \cite{ayoub2003eccentricity}:
\begin{equation}
    \mathcal{E}_1 =  \left\{  \frac{2\left[ (A-C)^2 + B^2 \right]^{1/2}}{(A+C)+\left[ (A-C)^2 + B^2 \right]^{1/2}} \right\}^{1/2}
\end{equation}
For $A = \vec{y}^2$, $B=-2(\vec{x}.\vec{y})$, and $C = \vec{x}^2$, this can be rewritten as:
\begin{equation}
    \mathcal{E}_1 = \left\{  \frac{2\left[ (\vec{x}^2-\vec{y}^2)^2 + 4(\vec{x}.\vec{y})^2 \right]^{1/2}}{(\vec{x}^2+\vec{y}^2)+\left[ (\vec{x}^2-\vec{y}^2)^2 + 4(\vec{x}.\vec{y})^2 \right]^{1/2}} \right\}^{1/2} \ .
\label{calc_E1}
\end{equation}
Note that $(\vec{x}^2-\vec{y}^2)^2 + 4(\vec{x}.\vec{y})^2$ can be represented in $U$ and $V$ as follows:
\begin{equation}
\begin{split}
    (\vec{x}^2-\vec{y}^2)^2 + 4(\vec{x}.\vec{y})^2 &= \left[ (\vec{x}^2+\vec{y}^2)^2 - 4 \vec{x}^2 \vec{y}^2\right] + 4 (\vec{x}.\vec{y})^2
    \\
    &=(\vec{x}^2+\vec{y}^2)^2 - 4 \left[ \vec{x}^2 \vec{y}^2 - (\vec{x}.\vec{y})^2 \right] 
    \\
    &= 4U^2 - 4V =  4(U^2-V) \ ,
\end{split}
\end{equation}
in which we have utilized Eq. \eqref{also_V} to go from the second- to the third-line. Plug this expression back to Eq. \eqref{calc_E1}, we get:
\begin{equation}
    \mathcal{E}_1 = \left[ \frac{2(U^2-V)^{1/2}}{U + (U^2-V)^{1/2}} \right]^{1/2} \ .
\end{equation}
The third-eccentricity $\mathcal{E}_3$ can be calculated from the first-eccentricity $\mathcal{E}_1$:
\begin{equation}
    \mathcal{E}_3 = \frac{\mathcal{E}_1}{(2-\mathcal{E}_1^2)^{1/2}} = (1-VU^{-2})^{1/4} = (1-W)^{1/4}  \ ,
\end{equation}
where $W=VU^{-2}$ as defined in Eq. \eqref{define_UW}.

The perimeter $\mathcal{L}$ of a general ellipse does not have a closed-form solution in terms of elementary functions \cite{chandrupatla2010perimeter}. Nevertheless, many simple approximations are available, and for simplicity let us consider a linear combination of $U^{1/2}$ and $V^{1/4}$, i.e.
\begin{equation}
    \mathcal{L} = \alpha U^{1/2} + \beta V^{1/4} \ , 
\end{equation}
in which the exponents here are chosen so that both terms scale linearly with length. he coefficients $\alpha$ and $\beta$ can then be fixed by requiring the approximation to be exact in two limiting cases: when the ellipse is maximally symmetric (a circle), and when it is maximally asymmetric (needle-like).
\begin{itemize}
    \item Eq. \eqref{fuzzy_sphere_points} generates an unit circle region on the transverse plane when e.g.
    $$(x_1,x_2,x_3,y_1,y_2,y_3)=(1,0,0,0,1,0) \ , $$
    in which $U=2$ and $V=1$. The perimeter of this unit circle is $\mathcal{L}=2\pi$, therefore $\alpha$ and $\beta$ should satisfy:
    \begin{equation}
    2\alpha + \beta = 1 \ .
    \label{eqA}
    \end{equation}
    \item Eq. \eqref{fuzzy_sphere_points} generates a needle-like region whose extent on the transverse plane is twice the unit length when e.g. 
    $$(x_1,x_2,x_3,y_1,y_2,y_3)=(1,0,0,0,0,0) \ , $$
    in which $U=1/2$ and $V=0$. The perimeter of this needle-like ellipse is $\mathcal{L}=4$, therefore $\alpha$ and $\beta$ should satisfy:
    \begin{equation}
    \alpha/\sqrt{2} = 4 \ .
    \label{eqB}
    \end{equation}
\end{itemize}
From Eq. \eqref{eqA} and Eq. \eqref{eqB}, we can solve them to obtain $\alpha = 4\sqrt{2}$ and $\beta = 2\pi - \alpha$ as in Eq. \eqref{LS}. It is remarkable that this crude approximation never deviates more than $4\%$ from the true value.

\subsection{The Gauge-Constraint \label{app:gauge_constraint}}

To show that $\hat{\vec{K}} \Psi(U,V)=0$ always holds, we just need to derive $\hat{K} U = \hat{K} V = 0$. Let us look at how the operator $\vec{x}\times \vec{\nabla}_{\vec{x}}$ acts on the SO$(3)\times$SO$(2)$ invariant variables, $U$:
\begin{equation}
\begin{split}
    \vec{x}\times \vec{\nabla}_{\vec{x}} U & = \vec{x}\times \left\{ \vec{\nabla}_{\vec{x}} \left[ \frac12 \left(\vec{x}^2+\vec{y}^2 \right) \right] \right\} = \vec{x} \times \left(\frac12 \vec{\nabla}_{\vec{x}} \vec{x}^2 + \frac12 \vec{\nabla}_{\vec{x}} \vec{y}^2 \right) 
    \\
    &= \vec{x} \times \left( \vec{x} + 0 \right) = 0 \ ,
\end{split}    
\end{equation}
and $V$:
\begin{equation}
\begin{split}
    \vec{x}\times \vec{\nabla}_{\vec{x}} V &= \vec{x}\times \left[ \vec{\nabla}_{\vec{x}} \left(\vec{x} \times \vec{y} \right)^2 \right] = \vec{x}\times \left[ 2 \left(\vec{x} \times \vec{y} \right) \cdot \vec{\nabla}_{\vec{x}} \left(\vec{x} \times \vec{y} \right) \right]
    \\
    &=\vec{x}\times \left\{ 2 \left(\vec{x} \times \vec{y} \right) \cdot \left[ \left( \vec{\nabla}_{\vec{x}} \times \vec{x} \right) \cdot \vec{y} - \vec{x} \cdot \left( \vec{\nabla}_{\vec{x}} \times \vec{y} \right) \right]  \right\}
    \\
    &=\vec{x}\times \left[ 2 \left(\vec{x} \times \vec{y} \right) \cdot \left( 0 \cdot \vec{y} - \vec{x} \cdot 0 \right)  \right] = 0 \ .
\end{split}    
\end{equation}
The same goes for the operator $\vec{y}\times \vec{\nabla}_{\vec{y}}$ and hence $\hat{\vec{K}}$, which is the sum of them.

\section{Geometric Expectation Values for Gaussian Distribution of $(\vec{x},\vec{y})$ \label{app:Gauss}}

\begin{figure*}[!htbp]
\includegraphics[width=0.66\textwidth]{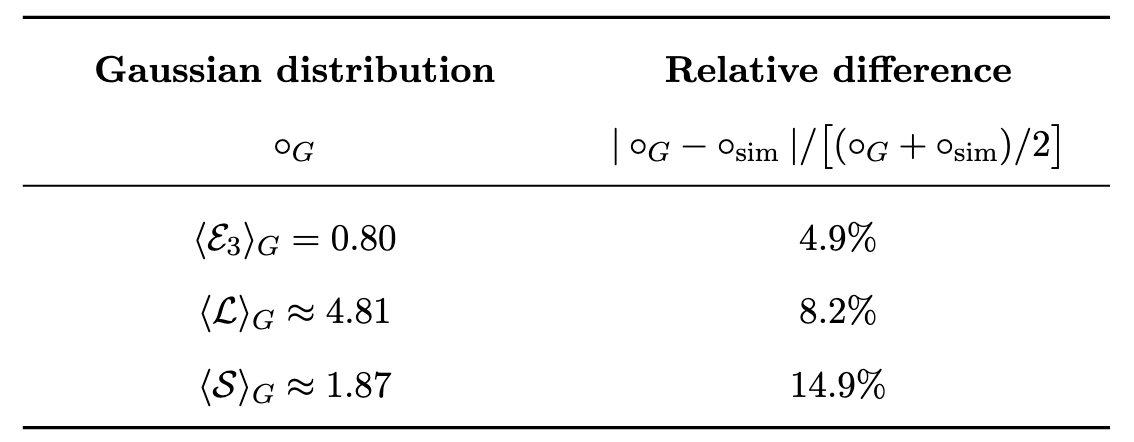}
\caption{\textbf{Comparison between estimates using a Gaussian distribution} (with $\langle \mathcal{A} \rangle_G = \langle \mathcal{A} \rangle_\text{sim}$) and those from the statistical simulation. The relative differences are small, but a few times larger than those in Fig. \ref{figT01}.}
\label{figT02}
\end{figure*}

We consider a baseline distribution $\mathcal{P}_G$ in which the values $(\vec{x},\vec{y})$ for the fuzzy sphere are randomly selected from a zero-mean Gaussian distribution with standard deviation (i.e. characteristic width) $\lambda$. Expressed in 
$UV$-space, this distribution can be written as:
\begin{equation}
    \mathbb{P}_G(U,V) = (2\lambda^6)^{-1} \exp\left( -U\lambda^{-2} \right) \ .
\end{equation}
Using this distribution instead of $\mathbb{P}_\Omega(U,V)$ in Eq. \eqref{calculate_expectation}, similar to what have been done in Section \ref{sec:theo_approx}, we can calculate the expectation area, eccentricity, perimeter, and shape parameter, which gives us the following results:
\begin{equation}
\begin{split}
    \langle \mathcal{A} \rangle_G &= \lambda^2 \times 4\pi \ , \ \langle \mathcal{E}_3 \rangle_G = 4/5 = 0.8 \ ,
    \\
    \langle \mathcal{L} \rangle_G &= \lambda \times \frac{3\sqrt{\pi} \left( \sqrt{2} + 2\pi \right)}{4} \approx  \lambda \times 10.23 \ , 
    \\
    \langle \mathcal{S} \rangle_G &= 1+\frac{4\sqrt{2} \left( \sqrt{2}+\pi \right)}{3\pi^2} \approx 1.870 \ .
\end{split}
\end{equation}
If we set $\lambda$ so that $\langle \mathcal{A} \rangle_G$ matches with $\langle \mathcal{A} \rangle_\text{sim} $, then $\lambda \approx 0.470$. We can then use this value for the calculation of the expectation perimeter, which yields $\langle \mathcal{L} \rangle_G \approx 4.812$. We compare these geometric values against the values reported in Fig. \ref{figT01} and find that, although the discrepancies are small, they still substantially exceed the offsets between the theoretical approximations and the simulation estimates -- see Fig. \ref{figT02}.

\section{Numerical Approach to Obtain Ground State Wavefunction \label{agent_based_sim}}

\subsection{Rescaling the Schr\"{o}dinger Equation \label{rescaling_Shro}}

Let us look at the following Hamiltonian, corresponding to $\kappa \rightarrow \kappa_*$ in Eq. \eqref{Hamiltonian_UV}:
\begin{equation}
\hat{H}\Big|_{\kappa \rightarrow \kappa_*} \equiv \hat{H}_{*} = -\frac12 \left( \vec{\nabla}_{\vec{x}}^2 + \vec{\nabla}_{\vec{y}}^2  \right) + \kappa_* V \ .
\end{equation}
Suppose $\Psi_*(U,V)$ is a normalized eigenfunction of $\hat{H}_{*}$ with corresponding eigenenergy $E_*$, i.e.
\begin{equation}
\hat{H}_{*} \Psi_*(U,V) = E_* \Psi_*(U,V) \ .
\label{Ham_simp}
\end{equation}
If we rescale the variables $\left( \vec{x}, \vec{y}\right) \rightarrow \lambda \left( \vec{x}, \vec{y}\right)$, which makes $\left( \vec{\nabla}_{\vec{x}}^2 + \vec{\nabla}_{\vec{y}}^2  \right) \rightarrow \lambda^{-2} \left( \vec{\nabla}_{\vec{x}}^2 + \vec{\nabla}_{\vec{y}}^2  \right)$, $U \rightarrow \lambda^2 U$, and $V \rightarrow \lambda^4 V$, then the right-hand side of Eq. \eqref{Ham_simp} becomes:
\begin{equation}
\hat{H}_{*} \Psi_*(U,V) \ \rightarrow \ \left[ -\frac12 \lambda^{-2} \left( \vec{\nabla}_{\vec{x}}^2 + \vec{\nabla}_{\vec{y}}^2  \right) + \lambda^4 \kappa_* V \right] \Psi_*(\lambda^2 U,\lambda^4 V)  \ .
\end{equation}
Take $\lambda =  (\kappa/\kappa_*)^{1/6}$, this equation becomes:
\begin{equation}
\begin{split}
(\kappa/\kappa_*)^{-1/3} \left[ -\frac12 \left( \vec{\nabla}_{\vec{x}}^2 + \vec{\nabla}_{\vec{y}}^2  \right) + \kappa V \right] \Psi_*\left[ (\kappa/\kappa_*)^{1/3} U,(\kappa/\kappa_*)^{2/3} V\right] &
\\
=  (\kappa/\kappa_*)^{-1/3} \hat{H} \Psi_*\left[ (\kappa/\kappa_*)^{1/3} U,(\kappa/\kappa_*)^{2/3} V\right] & \ .
\end{split}
\label{kappa_scale}
\end{equation}
This means the wave-function $(\kappa/\kappa_*) \Psi_*\left[ (\kappa/\kappa_*)^{1/3} U,(\kappa/\kappa_*)^{2/3} V\right]$ should be a normalized eigenfunction of $\hat{H}$ with corresponding eigenenergy $(\kappa/\kappa_*)^{1/3} E_*$.

\subsection{The Agent-Based Simulation \label{app:gsw}}

Here, we explain the details of the simulation done for Section \ref{sec:stat_sim}.

\subsubsection{Theoretical Basis for Estimating the Ground State with Fitness Landscape Simulation \label{Theo_basis_Sim}}

By simulating many random walkers on a fitness landscape and analyzing their stationary distribution, we can infer ground state properties of the corresponding quantum mechanical problem. The idea is proposed in \cite{ao2023schrodinger}, and let us briefly explain how it applies to the Hamiltonian given by Eq. \eqref{Hamiltonian_UV}. Consider a population of random walkers with effective diffusivity $D_0$ moving in a six-dimensional fitness landscape $(\vec{x},\vec{y})$, where the local growth rate is given by:
\begin{equation}
    R(\vec{x},\vec{y})\left[1-S(t)\right] \ \ \text{in which} \ \ R(\vec{x},\vec{y}) = R_0  \left[ 1-V(\vec{x},\vec{y}) \right] \ ,
\label{growth_profile}
\end{equation}
where the potential $V(\vec{x},\vec{y})$ is the same as in Eq. \eqref{Hamiltonian_UV}. Here $S(t)$ denotes the success of the total population at time $t$ \cite{phan2021doesn}, defined as:
\begin{equation}
    S(t) =  \frac{\int d^3\vec{x} d^3\vec{y} \ b (\vec{x},\vec{y},t)}{K} \ ,
\label{success}
\end{equation}
with $b(\vec{x},\vec{y},t)$ the population distribution and $K$ the carrying capacity. After sufficient time, the system reaches a stationary distribution $b_{st}(\vec{x},\vec{y})$ and a stationary success value $S_{st}$. In this steady state, $b_{st}$ satisfies:
\begin{equation}
    0 = D_0 \left( \vec{\nabla}_{\vec{x}}^2 + \vec{\nabla}_{\vec{y}}^2 \right) b_{st}(\vec{x},\vec{y})  + R(\vec{x},\vec{y}) (1-S_{st}) b_{st}(\vec{x},\vec{y}) \ .
\label{dist_st}
\end{equation}
Using Eq. \eqref{growth_profile} and rearranging this equation, we arrive at:
\begin{equation}
    \left[ -\frac12 \left( \vec{\nabla}_{\vec{x}}^2 + \vec{\nabla}_{\vec{y}}^2 \right) + \frac{R_0 \left( 1-S_{st}\right)}{2 D_0} V(\vec{x},\vec{y}) \right] b_{st}(\vec{x},\vec{y}) = \left[ \frac{R_0 \left( 1-S_{st}\right)}{2 D_0} \right] b_{st}(\vec{x},\vec{y}) \ .
\end{equation}
Define
\begin{equation}
    E_0 = \frac{R_0(1-S_{st})}{2D_0} \ ,
\label{define_E0}
\end{equation}
then by matching it with the Hamiltonian given by Eq. \eqref{Hamiltonian_UV}, we can see that $b_{st}(\vec{x},\vec{y})$ corresponds to the ground state wavefunction of the Hamiltonian $\hat{H}\Big|_{\kappa = E_0}$ with ground state energy $E_0$. Upon changing variables $(\vec{x},\vec{y}) \rightarrow (U,V)$ then following Appendix \ref{rescaling_Shro} in which we need to use $\kappa_*=E_*=E_0$ for Eq. \eqref{kappa_scale}, the ground state wavefunction and ground state energy for the Hamiltonian $\hat{H}$ -- as estimated with our simulation -- can be obtained from:
\begin{equation}
    \Psi_\text{sim}(U,V) \propto b_{st}\left[\left(\frac{\kappa}{E_0} \right)^{1/3}U,\left(\frac{\kappa}{E_0} \right)^{2/3}V \right] \ , \ E_\text{sim} = \left(\frac{\kappa}{E_0} \right)^{-1/3} E_0 \ .
\label{E_sim_to_all}
\end{equation}

\subsubsection{Fitness Landscape Simulation \label{Statistical_Sim}}

\begin{figure*}[!htbp]
\includegraphics[width=\textwidth]{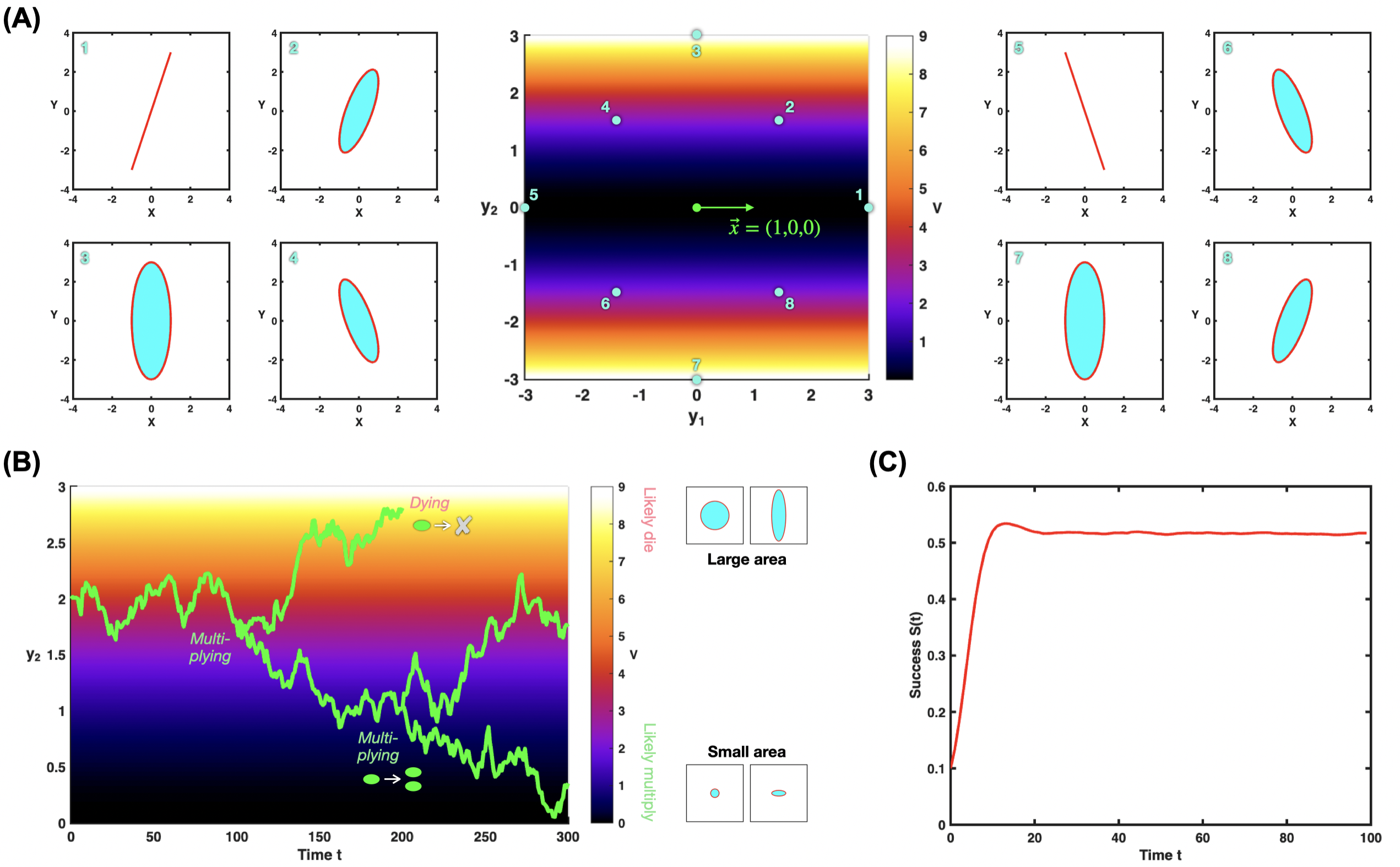}
\caption{\textbf{The simulation of many random-walkers on a fitness landscape.} We illustrate the model setup, walker dynamics, and how the population approaches the stationary state. \textbf{(A)} The potential $V(\vec{x},\vec{y})$, in the $y_1y_2$-plane where $y_3=0$ and $\vec{x}=(1,0,0)$. We further show how distinct positions in this six-dimensional space correspond to different fuzzy sphere projections. Note that some positions correspond to the same projection. \textbf{(B)} Consider a random-walker constrained on the $y2$-line with $(y_1,y_3)=(0,0)$ and $\vec{x}=(1,0,0)$. If it moves toward the center $y_2=0$, it is likely to multiply; if it moves away, it is likely to die. Here we show the trajectory $y_2(t)$ of a random walker, multiplication causes the trajectory to split into two, while death terminates it. \textbf{(C)} The change of success $S(t)$ with time indicate that this population of random walkers has reached the stationary state long before $t=50$.}
\label{fig04}
\end{figure*}

We use an agent-based simulation to investigate the population dynamics on a fitness landscape with a heterogeneous growth $R(\vec{r})$ described by Eq. \eqref{growth_profile}, in which each agent is specified by its location in a six-dimensional Euclidean space $\vec{r} \equiv (\vec{x},\vec{y})$ as follows:
\begin{equation}
\begin{split}
    r_1=x_1 \ , \ r_2=x_2 \ , \ r_3 = x_3 \ , \ r_4=y_1 \ , \ r_5 = y_2 \ , \ r_6 = y_3 & \ , 
    \\
    \text{and} \ \ V(\vec{r}) = (r_2 r_6 - r_3 r_5)^2 + (r_3 r_4 - r_1 r_6)^2 + (r_1 r_5 - r_2 r_4)^2 & \ . 
\end{split}
\label{xy_to_r_and_r_to_V}
\end{equation}
We discretize the time $t$ into evenly-pacing simulation time-steps, so that two consecutive steps are always $\Delta t$ apart. At each simulation step, the position of every agent in each direction $j\in \{ 1,2,3,4,5,6\}$ is updated according to a Wiener process:
\begin{equation}
r_j (t+\Delta t) = r_j(t) + \left( 2 D_0 \Delta t \right) \times \mathcal{N}(0,1) \ , 
\end{equation}
where $D_0$ is the effective diffusivity, and $\mathcal{N}(0,1)$ is a sample values from a Gaussian distribution of mean value $0$ and standard deviation $1$. Each agent also has a chance to multiply, die, or persist. These possibilities are controlled by a single value $p$:
\begin{equation}
p = R(\vec{r}) \left[1 - \frac{N(t)}{K} \right] \Delta t \ ,
\end{equation}
in which $K$ is the carrying capacity and $N(t)$ is the total number of agents at time $t$. The ratio $S(t)=N(t)/K$ is the success -- see Eq. \eqref{success}. A random number is generated uniformly between $[0,1]$, and if that number is larger than $|p|$ then the agent will multiply if $p>0$ and will die if $p<0$ (else, nothing will happen) at the end of the timestep. 

We summarize the information about our simulation in Fig. \ref{fig04}. The trajectory of a random-walker on this six-dimensional landscape corresponds to a history of fuzzy sphere projections, which we demonstrate in Supplementary Movie \textit{smovie01.mp4}.

In the simulation conducted for this work (with \textit{MatLab R2025a} \cite{MATLAB-R2025a}), we use a time-discretization $\Delta t = 10^{-2}$ and the total time of $T=10^2$. The other parameters are $D_0=1/18$, $R_0=1$, and $K=10^6$ (which serve as the cap for the number of agents). If a walker goes too far from the origin of the landscape, i.e. $|\vec{r}|>L_\text{IR}$ with $L_\text{IR}=10$, it will be eliminated from the simulation. Initially, we place $N_0=10^5$ walkers -- uniformly and randomly -- inside a six-dimensional hypercube of side $2L=2$ centered at the origin of the landscape. From the observed behavior of $S(t)$, the population seems to reach the stationary state long before $t=50$ (see Fig. \ref{fig04}C), so we decide to use all random walker positions taken at different times $t\in \{51, 52, 53, ..., 100\}$ to estimate the stationary distribution and stationary success on the fitness landscape. For example, we obtain $S_{st} = 0.516 \pm 0.003$, within $95\%$ confident. Since the relative uncertainties of all simulation estimates are very small compared with their value deviations from the theoretical approximations, we omit them for convenience. The energy $E_0$, given by Eq. \eqref{define_E0}, is estimated to be $4.35$.

To obtain the probability distribution $\mathbb{P}_\Omega (U,V) = |\Psi_{\Omega}(U,V)|^2$ as in Eq. \eqref{probdist}, we first convert the six-dimensional positions $\vec{r}=(\vec{x},\vec{y})$ into $UV$-space, i.e. 
\begin{equation}
    U(\vec{r}) = \frac12 \left( r_1^2 + r_2^2 + r_3^2 + r_4^2 + r_5^2 + r_6^2 \right)
\end{equation}
which follows from Eq. \eqref{define_UW} and $V(\vec{r})$ with Eq. \eqref{xy_to_r_and_r_to_V}. Then we can rescale $U$ and $V$ as in Eq. \eqref{E_sim_to_all} since $E_\text{sim}$ is determined after knowing $S_{st}$, so that the distribution in this rescaled $UV$-space estimate the ground state wavefunction $\Psi_\Omega(U,V)$. We use a $k$-nearest-neighbor (k-NN) estimator \cite{loftsgaarden1965nonparametric} at each data point to obtain $\Psi_\Omega(U,V)$, then square it to get $\mathbb{P}_\Omega(U,V)$ (taking the absolute value of $\Psi_\Omega(U,V)$ is not necessary since it is always non-negative). Using k-NN gives agent-wise estimates, so downstream estimations -- in this work, these are for the area $\mathcal{A}$, the third-eccentricity $\mathcal{E}_3$, the perimeter $\mathcal{L}$, and the shape parameter $\mathcal{S}$, using Eq. \eqref{AE} and Eq. \eqref{LS} -- are defined per agent, which reduces estimation uncertainty since dense regions use many nearby points (lower variance), while sparse areas automatically use a wider neighborhood to tame noise. Overall, this makes k-NN approach a better choice than the conventional binning approach for our statistical analysis.

The code for these analyses is available from the corresponding author on request.

\bibliography{main}
\bibliographystyle{apsrev4-2}

\end{document}